\documentclass[aps,twocolumn,pre,nofootinbib]{revtex4}
\usepackage{graphicx}
\newcommand{\MCtwo}{Microtechnology and Nanoscience, MC2, 
Chalmers University of Technology, SE-412 96 G{\"o}teborg, Sweden}

\newcommand{\vdWDF}{{\mbox{\scriptsize vdW-DF}}}

\newcommand{\tot}{{\mbox{\scriptsize tot}}}

\newcommand{\MeOH}{{\mbox{\scriptsize MeOH}}}
\newcommand{\graphene}{{\mbox{\scriptsize graphene}}}
\newcommand{\MeOHgraphene}{{\mbox{\scriptsize MeOH on graphene}}}

\begin{document}

\title{Methanol adsorption on graphene}

\author{Elsebeth Schr{\"o}der}\email{schroder@chalmers.se}\affiliation{\MCtwo}

\begin{abstract}
The adsorption energies and orientation of methanol on graphene are 
determined from first-principles density functional calculations. 
We employ the well-tested vdW-DF method that seamlessly includes
dispersion interactions with all of the more close-ranged interactions 
that result in bonds like the covalent and hydrogen bonds. 
The adsorption of a single methanol molecule and small methanol 
clusters on graphene are studied at various coverages.
Adsorption in clusters or at high coverages (less than a monolayer)
is found to be preferable, with the methanol C-O axis approximately
parallel to the plane of graphene.
The adsorption energies calculated with vdW-DF are compared with 
previous DFT-D and MP2-based calculations for single methanol
adsorption on flakes of graphene (polycyclic aromatic hydrocarbons). 
For the high coverage adsorption energies we also find reasonably 
good agreement with previous desorption measurements. 
\end{abstract}

\date{March 14, 2013}

\maketitle

%%%%%%%%%%%%%%%%%%%%%%%%%%%%%%%%%%%%%%%%%%%%%%%%%%%%%%%%%%%%%%%%%%%

\section{Introduction}

Methanol (CH$_3$OH) is the simplest of the alcohols, and it is 
used, e.g., as a solvent, an alternative fuel, and as a source for producing 
other chemicals. 
Methanol is the second most abundant
organic molecule in the atmosphere next after methane (CH$_4$),
and along with other insoluble aerosol particles methanol is 
believed to play a role in the formation of ice in the atmosphere, 
as discussed and modeled e.g.\ in Refs.\ \cite{kong,thomson}.
Methanol is also found 
in the interstellar medium, as methanol ice dust grains.
The graphite surface is a suitable model for dust in the interstellar 
medium for such studies \cite{bolina,Wolff}.

In this paper we calculate by first-principles density functional theory 
(DFT) the adsorption energy of methanol on graphite at various degrees 
of coverage (less than one molecular monolayer), and we determine the 
distance from and the optimal angle of the methanol molecule C-O axis 
with the plane of graphene. For the DFT calculations we
use the method vdW-DF \cite{Dion,Thonhauser}.

Previously, the adsorption energies of methanol from graphene or 
flakes of graphene were calculated \cite{pankewitz} by the 
semi-empirical theory method DFT-D \cite{grimme}.
In the study of Ref.\ \cite{pankewitz} the adsorption of methanol
on to very small flakes of graphene---benzene and coronene---was 
also calculated by the higher-accuracy correlation method 
second-order M{\o}ller-Plesset perturbation theory (MP2).
Desorption energies from highly oriented pyrolythic graphite (HOPG) 
or from various sizes of single-walled
carbon nanotubes (SWCNTs) have also previously been measured in 
a number of desorption experiments \cite{hertel,bolina,burghaus}.   

The purpose of this study is to determine basic information about 
adsorbed methanol on graphene, such as the optimal orientation,
the interaction (adsorption) energy at various coverages and distances from 
graphene, and to make these available as input to and/or for fine-tuning of
molecular dynamics simulations of the methanol adsorption process. 

In the following, we first introduce the methanol-graphene system, the vdW-DF 
method, and the set up of our calculations. Next, we describe our results both 
at low and high coverage of methanol on graphene and then discuss the relation to 
the theory results of Ref.\ \cite{pankewitz} and some of the available 
experimental results.

\section{Materials and Methods}

On some surfaces methanol chemisorbs. When this is the case, traditional 
semilocal DFT methods, based on the generalized gradient expansion (GGA), 
may suffice for describing the adsorption. 
For instance, this is the case on the oxide surfaces
$\alpha$-Al$_2$O$_3$(0001) and $\alpha$-Cr$_2$O$_3$(0001) on which 
we previously 
studied methanol adsorption \cite{MeOHAl2O3,MeOHCr2O3}. 
However, on graphene a number of small molecules phys\-isorb, or at least 
owe a significant part of their adsorption energy to the dispersion 
interaction. Then GGA methods are inadequate.
 
We here use the vdW-DF method \cite{Dion,Thonhauser}. It includes the 
van der Waals (vdW) interactions (also termed the London dispersion 
interactions), 
that are especially important on intermediate to long ranges, along with 
all the 
traits of GGA for short-range interactions. Thus, vdW-DF delivers a 
description of the system that takes care of both the vdW 
interaction between the fragments (and within the fragments) and the 
short-ranged interaction within the molecules, like the covalent bonding, 
hydrogen bonding, possible ionic interactions etc., all from first principles.

Over the past few years, our group has carried out a series of 
physisorption studies of relatively small molecules on graphene: 
n-alkanes (of length 1 to 10 C atoms) \cite{alkanes}, 
phenol \cite{phenol}, 
small polycyclic aromatic hydrocarbons (PAHs) \cite{naphthalene,PAHdimers}, 
trihalomethanes \cite{chloroform},
adenine \cite{adenine}, 
and with somewhat different computational details, all of the five 
nucleobases of DNA and RNA \cite{nucleobasesgraphene}.  
General considerations of use of the vdW-DF method for such systems, 
as well as further method discussions, can be found in the
Refs.~\cite{alkanes,phenol,naphthalene,PAHdimers,chloroform,adenine} 
mentioned above.

We here use the DFT program GPAW \cite{gpaw} 
with a fast-Fourier-transform implementation of 
vdW-DF \cite{Dion,Thonhauser,soler}. Pre- and postprocessing is carried out
in the python environment ASE \cite{ASE}.

\begin{table*}
\caption{\label{tab:energies} Theory data for adsorption of methanol
on graphene. Included is the adsorption energy $E_a$, 
the distance of the methanol O atom from graphene, $d_{O}$,
the adsorption configuration 
(C-O axis approximately parallel with the plane of
graphene, or C-O axis approximately perpendicular to the plane of graphene
with the O atom pointing up or down), molecular coverage,
and the orthogonal unit cell used in the calculations, given in 
units of $a_g=\sqrt{3} \,a_0$ with $a_0=1.43$ {\AA}. All unit cells are 
19 {\AA} in the direction perpendicular to the plane of graphene.
The coverages in our calculations are estimated from the approximate
molecular area of methanol on graphene 17.6 {\AA}$^2$ determined from
x-ray diffraction studies in Ref.\ \cite{morishige}.
}

\begin{tabular}{llccccc}
 \hline\hline
                       & Structure& Unit cell            &Coverage&\multicolumn{2}{c}{$E_a$}&$d_{O}$ \\
                       &          &                     &[ML]&[kJ/mol]&[meV]&[{\AA}] \\
  \hline
  \multicolumn{6}{l}{\textit{Theory, vdW-DF (our calculations)}}\\
  single molecule      & parallel & $3\sqrt{3}\times 5$ &0.11& 20.6 & 214 & 3.33 \\
                       & parallel & $3\sqrt{3}\times 4$ &0.14& 20.7 & 215 & 3.33 \\
                       & parallel & $3\sqrt{3}\times 3$ &0.18& 21.1 & 219 & 3.33 \\
                       & O down   & $3\sqrt{3}\times 3$ &0.18& 14.6 & 151 & 3.20 \\
                       & O up     & $3\sqrt{3}\times 3$ &0.18& 15.5 & 160 & 4.87 \\
                       & parallel & $1\sqrt{3}\times 2$ &0.83& 30.5 & 316 & 3.55 \\
  three-cluster        & parallel & $3\sqrt{3}\times 3$ &0.55& 30.4 & 315 & 3.31--3.54 \\
  five-cluster         & parallel & $3\sqrt{3}\times 3$ &0.92& 34.9 & 361 & 3.35--4.50 \\
  \multicolumn{6}{l}{\textit{Theory, DFT-D and MP2-based (Pankewitz and Klopper)$^a$}}\\
  single molecule, DFT-D with BP86 &parallel&   benzene && 14.8 & & 3.35 \\
  single molecule, DFT-D with BP86 &parallel& coronene  && 18.7 & & 3.32 \\
  single molecule, DFT-D with BP86 &parallel& 112-C PAH && 20.0 & & 3.32 \\
  single molecule, DFT-D with BP86$^{b}$ &O up& coronene  &&$\sim$11&&$\sim$4.8 \\
  single molecule, SCS-MP2 with PB86 &parallel& benzene && 13.8 & & $\sim$3.4 \\
  single molecule, SCS-MP2 with PB86 &parallel& coronene&& 18.3 & & $\sim$3.3 \\
\hline
\multicolumn{7}{l}{${}^a$Orbital-based calculations with a TZVP basis, Ref.~\protect\cite{pankewitz}.}\\
\multicolumn{7}{l}{${}^b$Energy estimated from Figure 8 of 
Ref.~\protect\cite{pankewitz}.}%\\

\end{tabular}
\end{table*}

\begin{figure}[t]
\begin{center}
\includegraphics[width=0.4\textwidth]{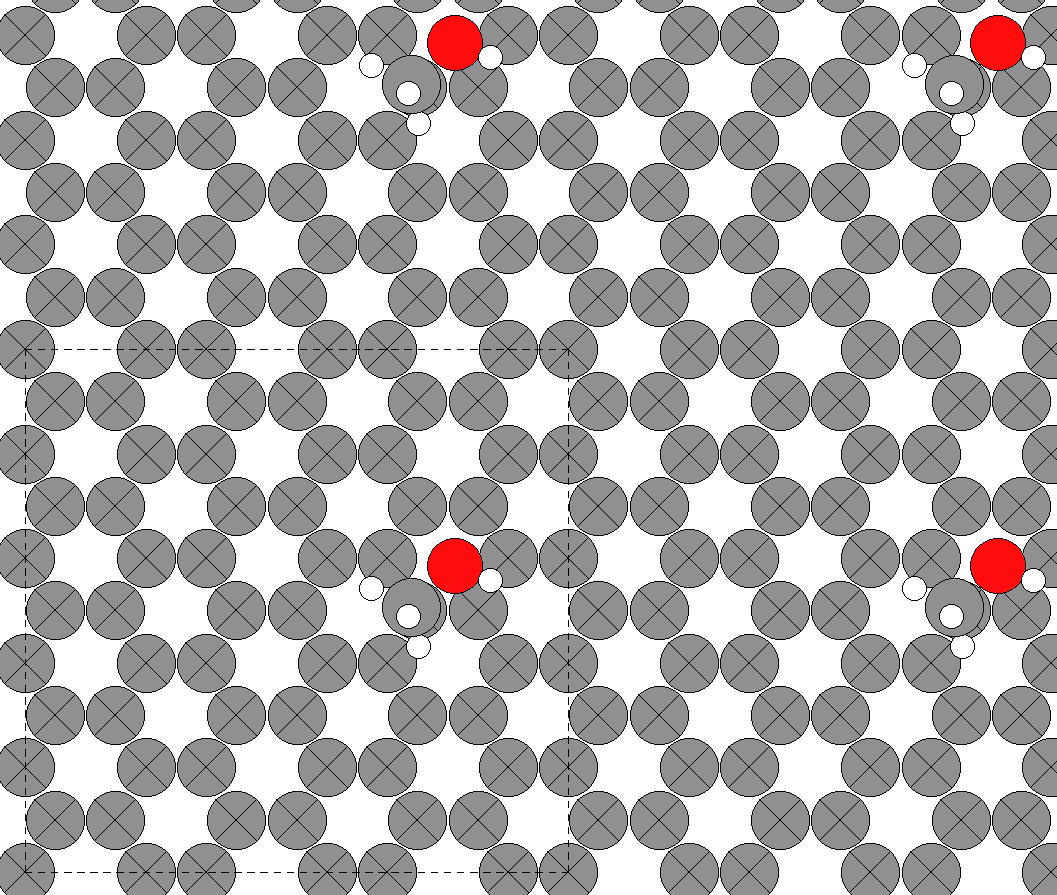}
\\
\includegraphics[width=0.4\textwidth]{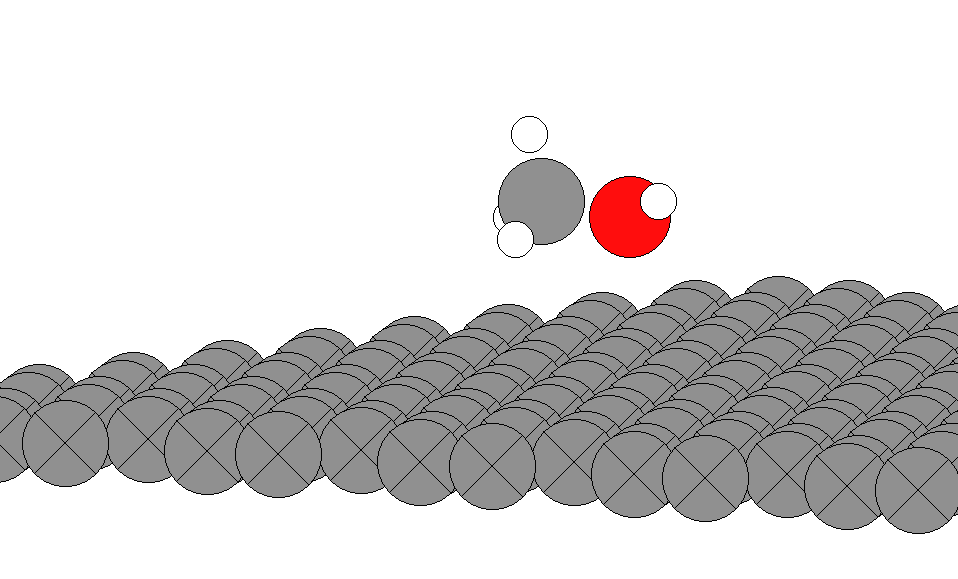}
\caption{\label{fig:config1}Schematic view of a single
methanol molecule adsorbed on graphene in the $3\sqrt{3}\,a_g\times 5a_g \times 19$ {\AA} 
periodically repeated unit cell. The configuration with the C-O axis
approximately parallel with graphene is shown. Gray circles with a cross
are graphene C atoms. Other gray/red/small white circles are the methanol C, O and H atoms.
In the top panel the unit cell is outlined by
thin broken lines.  }
\end{center}
\end{figure}

\begin{figure}[t]
\begin{center}
\includegraphics[width=0.4\textwidth]{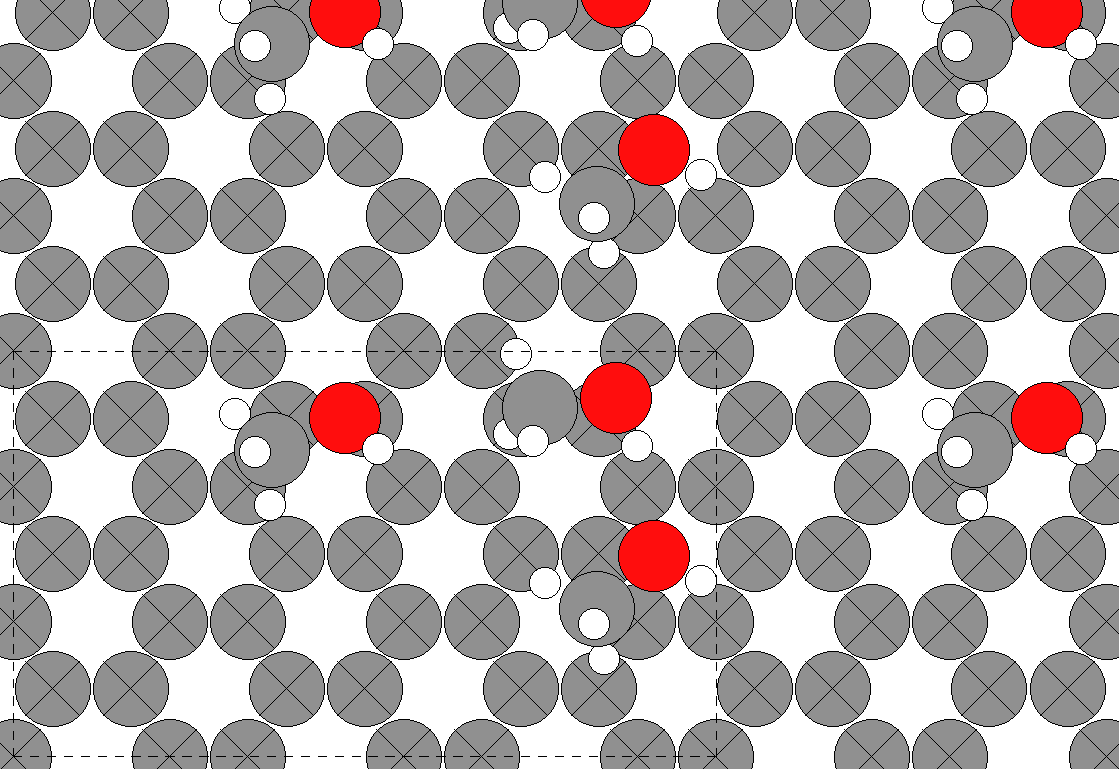}
\\[1em]
\includegraphics[width=0.4\textwidth]{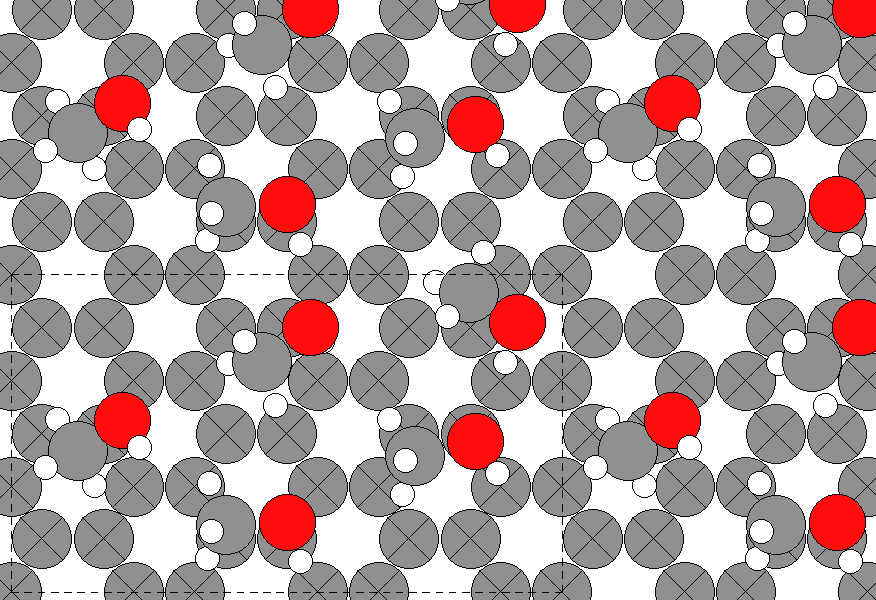}
\caption{\label{fig:config2}
Three- and five-molecule clusters in the 
$3\sqrt{3}\,a_g\times 3a_g\times 19$ {\AA} unit cell.  
 }
\end{center}
\end{figure}

We use periodic orthorhombic unit cells as detailed in 
Table \ref{tab:energies}, with 8--60 graphene C-atoms per unit cell. 
The unit cell side lengths in the plane of graphene 
range from 4.29 to 12.87~{\AA} with one, three or five methanol molecules
per unit cell, as illustrated for one molecule in Figure \ref{fig:config1} and
for three and five molecules in Figure \ref{fig:config2}. 

The wave functions are sampled on a regular grid with points 0.12 {\AA} apart 
(the charge density is sampled at points half that distance apart) to obtain
sufficient accuracy for high-quality results from the vdW-DF 
calculations \cite{alkanes,Kintercalation}.
The reciprocal space $k$-point sampling is $2\times 2\times 1$ for the largest
and $6\times 8\times 1$ for the smallest unit cell, except for the
calculation of isolated methanol where only the $\Gamma$-point is used.

In all calculations the atomic positions are allowed to relax.
We use the molecular-dynamics optimization method ``fast inertial
relaxation engine" (FIRE) \cite{fire} and require that the
remaining force on each atom has a size of less than 0.01 eV/{\AA}.

We determine the adsorption energy $E_a$ as the difference in total 
energies of the full system $E^\vdWDF_{\tot,\MeOHgraphene}$ 
and each fragment isolated,
\begin{eqnarray}
E_a &=& E^\vdWDF_{\tot,\MeOHgraphene}  \nonumber \\
&&{}- E^\vdWDF_{\tot,\graphene} - E^\vdWDF_{\tot,\MeOH} \, .
\label{eq:Ea}
\end{eqnarray}
Here the first two terms are found using the unit cell size of the full 
system (Table \ref{tab:energies}), whereas the last term is calculated
in a $3\sqrt{3}\,a_g\times 5a_g \times 19$ {\AA} unit cell with only
$\Gamma$-point sampling. For the three- or five-molecule cluster
calculations, equation (\ref{eq:Ea}) is changed appropriately to give the 
adsorption energy $E_a$ per adsorbate (in eV) or per mol (in kJ). 

The data points of the potential energy-curve in Figure \ref{fig:Ez} are 
obtained with a slightly longer unit cell than the other calculations: 
because we need to calculate
the methanol-graphene interaction at up to relatively large separations
(11 {\AA}) the unit cell height is increased to 
23 {\AA}, all other settings remaining the same.
 
\section{Results and Discussion}

\begin{figure}[h]
\begin{center}
\includegraphics[width=0.5\textwidth]{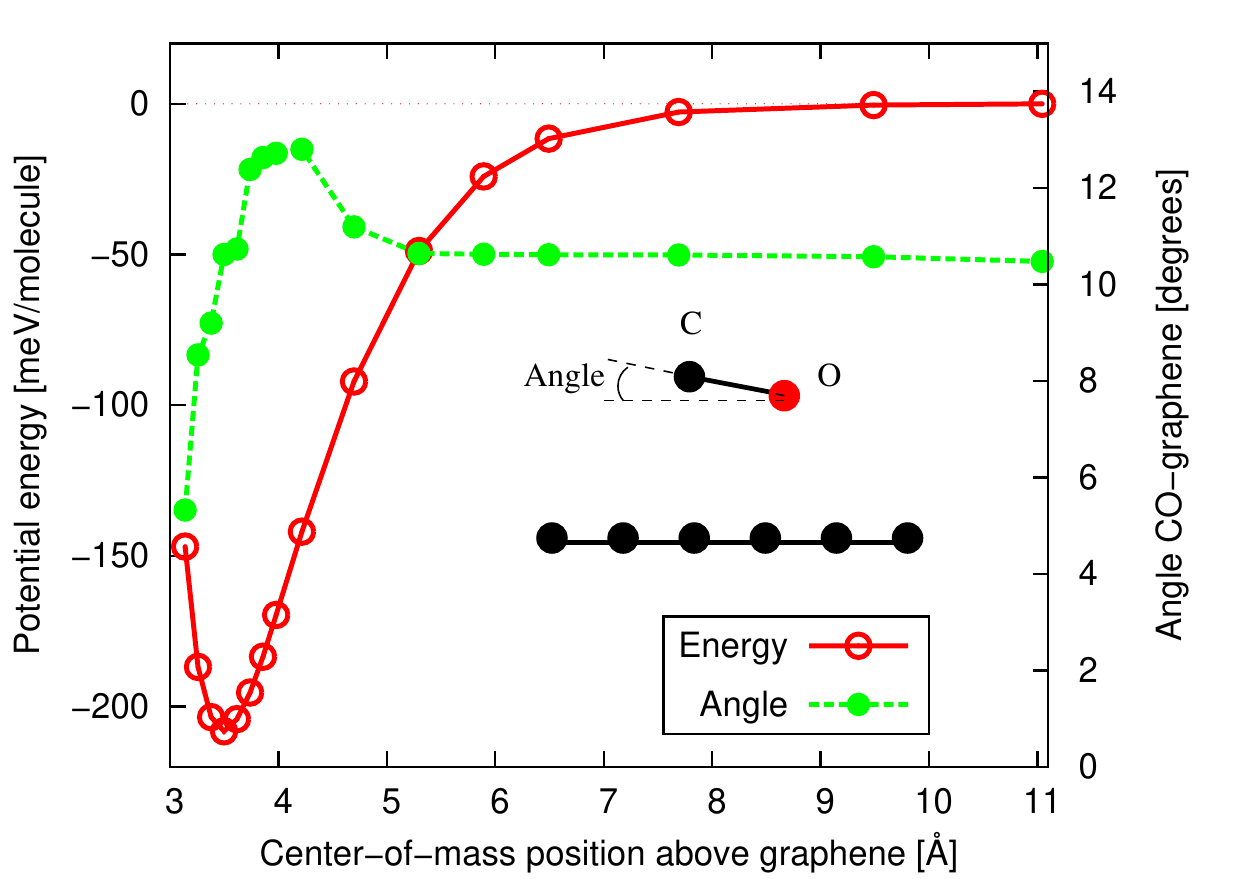}
\caption{\label{fig:Ez}Potential energy of a single molecules of methanol
on graphene at various distances from graphene (open circles, left graph axis).
In each calculated point of the curve the atoms of the methanol molecule
are allowed to relax according to the forces on the
atoms, with the constraint that the center-of-mass position is fixed.
The angle that the C-O axis makes with the graphene plane is shown in
the curve with solid circles (right graph axis). The insert defines the
angle.}
\end{center}
\end{figure}

In Table \ref{tab:energies} we list the adsorption energies $E_a$
for the various methanol-graphene systems that we consider. 
Also shown for each calculation is the coverage of methanol on graphene
in units of molecular monolayers (ML), derived from the estimated
area per molecule 17.6 {\AA}$^2$ at 1 ML obtained by Morishige et al.\ \cite{morishige}
from x-ray diffraction studies of a methanol film on graphite.

In the $3\sqrt{3}\,a_g\times 3a_g$ unit cell we test starting
the calculations with methanol oriented such that the C-O axis
is either parallel or perpendicular to graphene (with the O atom
pointing away from or towards graphene). As evident from Table \ref{tab:energies},
the almost-parallel orientation, after relaxation of 
the atomic positions, is energetically more favorable than
the two perpendicular orientations (``O up" and ``O down"). 
While the parallel orientation gives rise to a 219 meV (21.1 kJ/mol) 
binding, the 
perpendicular orientations only bind with 160 and 151 meV (15.5 and 14.6 kJ/mol).
For the remaining calculations we therefore start methanol
oriented parallel to graphene.\footnote{We did check starting with
the ``O down"
orientation in the small $1\sqrt{3}\,a_g\times 2a_g $
unit cell, but after full
relaxation of the atomic positions methanol ended up with an 
orientation parallel to graphene.}

Ignoring the $E_a$ of the two perpendicular orientations
(``O up" and ``O down") we see that $E_a$ grows with increasing
coverage, from 214 meV at 0.11 ML to 361 meV at 0.92 ML, the exception 
being the single molecule in the $1\sqrt{3}\,a_g\times 2a_g $
unit cell.
We also find that the methanol-graphene distance, here measured as the distance
to  O in methanol, $d_O$, slightly increases with coverage, although 
the trend is not clear for the cluster calculations.

As seen by the growth in $E_a$ with coverage, the methanol molecules
interact attractively, across unit cell boundaries (due to periodic 
boundary conditions) and for the cluster calculations also within the 
unit cell. The nearest-neighbor adsorbate-adsorbate distance in our 
single-molecule calculations varies from 12.4 {\AA} in the largest unit cell
to 4.3 {\AA} in the smallest unit cell. In the cluster calculations 
the O-to-O nearest-neighbor separation is 3.0 {\AA} in the three-adsorbate 
cluster (one such interaction per unit cell) and 2.9 {\AA} 
in the five-adsorbate cluster (two such nearest neighbors per unit cell).

It should be noted that  the cluster calculations are started with methanol
distributed at ``reasonable" inter-molecular separations, not specifically
in any expected cluster-configuration, and the relative orientations
are not optimized for the orientations at e.g.\ a full ML \cite{morishige}.

It is possible to partition the adsorption energy into the contributions 
from the substrate-adsorbate interaction and the adsorbate-adsorbate 
interaction. The adsorbate-adsorbate interaction energy is found from 
the total energy of the system with the graphene substrate removed, all
other atom positions unchanged, and subtracting the total energy of an 
isolated molecule (times three or five for the clusters). 

For the $3\sqrt{3}\,a_g\times 4 a_g$ unit cell this attractive interaction 
across cell boundaries amounts to a mere 2 meV per molecule (0.2 kJ/mol), 
increases to 7 meV (0.7 kJ/mol) for the $3\sqrt{3}\,a_g\times 3a_g$ unit cell,
and 125 meV (12.1 kJ/mol) for the much more dense phase in the 
$1\sqrt{3}\,a_g\times 2a_g$ unit cell.
For the three-cluster system the sum of adsorbate-adsorbate interactions 
per unit cell is 366 meV, thus in average 122 meV per molecule (11.8 kJ/mol). 
However, as noted above and seen also in the top panel of 
Figure \ref{fig:config2}, one pair of molecules per unit cell is closer 
together than the other pair interactions, and this pair interaction
is thus
expected to dominate the sum of interaction energies. This explains 
why the three-cluster system yields an adsorbate-adsorbate energy  
at only 0.55 ML that is almost the same as the evenly distributed 
molecules in the $1\sqrt{3}\,a_g\times 2a_g$ unit cell at 0.83 ML.

In the five-cluster calculations, the adsorbates are in reality almost 
uniformly distributed within the unit cell, and can hardly be considered a 
``cluster". The coverage is close to a full ML and all molecules are 
relatively close to each other, although two O-O separations
stand out as being smaller. 
We find the sum of adsorbate-adsorbate interactions per unit cell
909 meV, which yields an average 182 meV per adsorbate (17.5 kJ/mol),
which is the largest adsorbate-adsorbate interaction energy of
this study.  

Thus the methanol-methanol interactions are important as they contribute 
a large fraction of the adsorption energy, but the methanol-graphene  
interaction is stronger. As discussed e.g.\ in Ref.\ \cite{bolina} 
this facilitates the formation of a full methanol monolayer prior to 
forming multilayers, because the energy gain for the methanol molecule 
is larger when binding to graphene than to other methanol molecules.

In Figure \ref{fig:Ez} the potential-energy curve of various graphene-methanol
separations is shown. In this figure, each data point (circle) is found
by keeping the center of mass of methanol relative to the
plane of graphene at the distance $d_{cm}$, shown on the bottom axis
of the figure.
The atoms of the molecule are allowed to move in all directions, as long
as $d_{cm}$ remains unchanged. Thus, the orientation of the C-O axis
changes with distance from graphene, as shown by the filled circles
of Figure \ref{fig:Ez}.
When methanol is squeezed close to graphene ($d_{cm}\approx 3.2$~{\AA})
the angle is smallest (the orientation is closest to being parallel),
as a way for methanol to ``avoid" a too close contact to graphene of any
of its atoms.
At the adsorption distance, the angle is approximately 10.6$^\circ$.
As the fragments are further separated, the angle grows a bit until 
the distance $d_{cm}\approx 5$ {\AA} where the vdW interaction is too weak
to change the angle from the initial angle 10.6$^\circ$ (each calculation
is started with methanol in the adsorption configuration, translated
towards or away from graphene).

Pankewitz and Klopper \cite{pankewitz} carried out non-periodic DFT-D 
calculations of methanol adsorbed on SWCNTs and PAH-models of graphene of size
from benzene up to a PAH with 112 C atoms.
Although the DFT-D calculations are semi-empirical and thus can be 
less accurate (depending on the choice of empirical parameters for each 
type of calculation) the adsorption energies on PAH
(Table \ref{tab:energies}) agree reasonably with the present results,
when the smaller substrate size in the DFT-D calculations
(due to lack of periodicity) is taken into 
account \cite{grimmePAHsize,adenine}.
For adsorption on to benzene and the PAH coronene (24 C) they also
carried out spin-component scaled MP2 (SCS-MP2).
Their SCS-MP2 result for methanol on coronene is in good agreement with the 
present results: 
{From} the DFT-D calculations we can estimate that approximately 
1.3 kJ/mol of the methanol-substrate interaction is missing on coronene 
compared to graphene, or a rather large (112-C) PAH molecule. 
Their SCS-MP2 coronene result of 18.3 kJ/mol should therefore probably 
be corrected to $\sim$19.6 kJ/mol for a single methanol molecule adsorption 
on graphene. For our largest unit cell we find 20.6 kJ/mol and less 
than a 0.1 kJ/mol correction for the periodicity. Thus
our calculated energy for single methanol molecule adsorption deviates 
less than 1 kJ/mol (or 5\%) from the estimate of the size-modified 
SCS-MP2 results. 
The distance of methanol O from graphene, $d_O$, turns out identical 
in the two calculations. 

Interestingly, in the DFT-D calculations Pankewitz and Klopper find
a second, much weaker local minimum with the methanol O atom pointing 
away from coronene, much like our ``O-up" configuration. 
For this configuration the energies and 
substrate-to-O distances are very similar,
with 15.1 kJ/mol at 4.83 {\AA} in the vdW-DF calculations and 
approximately 11 kJ/mol at 4.8 {\AA} in the DFT-D calculations. 
In the binding energy curve of Figure \ref{fig:Ez} we do see a change in 
methanol angle with graphene as the distance is varied, but at the 
4.5--5 {\AA} center-of-mass
distance from graphene the interactions are probably too weak 
for the computational relaxation procedure to rotate the initially 
almost-parallel molecule to obtain the O-up structure.

Although we did no effort in fitting a full monolayer of methanol
on to graphene, it is still of interest to compare our high-coverage
results with other calculations of closely packed methanol molecules.
Boyd and Boyd \cite{boyd} used DFT with B3LYP at various basis set
levels to calculate the binding energies and structures of 
(free-floating) methanol clusters of up to 14 molecules.
They expect the inter-molecular interaction to be dominated by
the hydrogen bonds, for which B3LYP behaves reasonably. They find
the binding energy in the optimal clusters to be 27 kJ/mol, which is
larger than our largest molecule-molecule energy 17.5 kJ/mol
(for the 5-molecule cluster), but then, in our calculations 
there is still room for more molecules,   
and the molecules are constrained by the adsorption to graphene 
to form interactions only in two dimensions. Our smallest O-O distance
of 2.9 {\AA} (in the five-molecule cluster) compares well with
the optimal O-O distance 2.77 {\AA} in the trimer calculations
of Ref.\ \cite{boyd}.

A number of experiments of methanol desorption from HOPG
or SWCNTs have previously been carried out \cite{hertel,bolina,burghaus}. 
Although the desorption energies vary between the experiments, 
the energies for 1 ML coverage or less all fall in the range 28--51 kJ/mol
($48\pm3$ kJ/mol at 1 ML \cite{hertel}, 
33--48 kJ/mol at $< 1$ ML \cite{bolina}, 
28 kJ/mol at $< 1$ ML \cite{burghaus}), 
with a tendency to increase with increasing
coverage. The desorption energy range is in reasonable agreement with our results, 
that are in the range 20--35 kJ/mol (absent the less favorable 
methanol orientations), with increasing adsorption energy for increasing
coverage (Table \ref{tab:energies}). 
None of the mentioned experiments measure the distance of methanol 
from the substrate, nor the orientation of methanol. 
Further comparison to experiments that are presently in progress 
will appear in a forthcoming publication \cite{pettersson}.  

\section{Conclusions}

By use of the first-principles vdW-DF method we calculate adsorption energies and
determine adsorption geometries of methanol on graphene. Our results 
are in reasonable agreement with other available calculations and experiments. 
This suggests that the data obtained here may be used as input 
parameters to or tests of results from calculations and models
that (unlike DFT) rely on external information,
either from experiment or from (preferentially) first-principles calculations.
This could, for example, be models
that are on larger length scales or with 
time dependency, such as molecular dynamics calculations.

\acknowledgments
I thank Nikola Markovic, Chalmers, and Jan B.C.\ Pettersson, G\"oteborg
University, for introduction to this problem and for discussions.
The work was partially funded by the Swedish Research Council (VR) 
and the Chalmers Areas of Advances Nano and Materials. 
The computations were performed on resources provided by the Swedish 
National Infrastructure for Computing (SNIC) at C3SE. 

%%%%%%%%%%%%%%%%%%%%%%%%%%%%%%%%%%%%%%%%%%%%%%%%%%%%%%%%%%%%%%%%%%%%%%%%%%%%%


\begin{thebibliography}{99}
\bibitem{kong}
X. Kong, P.U. Andersson, E.S. Thomson, and J.B.C. Pettersson,
J. Phys. Chem. C \textbf{116}, 8964 (2012).

\bibitem{thomson}
E.S. Thomson, X. Kong, P.U. Andersson, N. Markovic, and J.B.C. Pettersson,
J. Phys. Chem. Lett. \textbf{2}, 2174 (2011).

\bibitem{bolina}
A.S. Bolina, A.J. Wolff, and W.A. Brown,
J. Chem. Phys. \textbf{122}, 044713 (2005).

\bibitem{Wolff}
A.J. Wolff, C. Carlstedt, and W.A. Brown,
J. Phys. Chem. C \textbf{111}, 5990 (2007).

\bibitem{Dion}
M. Dion, H. Rydberg, E. Schr\"oder, D.C. Langreth, and B.I. Lundqvist,
Phys. Rev. Lett. \textbf{92}, 246401 (2004); \textbf{95}, 109902(E) (2005).

\bibitem{Thonhauser}
T. Thonhauser, V.R. Cooper, S. Li, A. Puzder, P. Hyldgaard, and D.C. Langreth,
Phys. Rev. B \textbf{76}, 125112 (2007).

\bibitem{pankewitz}
T. Pankewitz and W. Klopper,
J. Phys. Chem. C \textbf{111}, 18917 (2007).

\bibitem{grimme}
S. Grimme, J. Comput. Chem. \textbf{25}, 1463 (2004).

\bibitem{burghaus}
U. Burghaus, D. Bye, K. Cosert, J. Goering, A. Guerard, E. Kadossov,
E.Lee, Y Nadoyama, N. Richter, E. Schaefer, J. Smith, D. Ulness,
and B. Wymore,
Chem. Phys. Lett. \textbf{442}, 344 (2007).

\bibitem{hertel}
H. Ulbricht, R. Zacharia, N. Cindir, and T. Hertel,
Carbon \textbf{44}, 2931 (2006).

\bibitem{MeOHAl2O3}
{\O}. Borck and E. Schr\"oder,
J. Phys.: Cond. Matter \textbf{18}, 1 (2006).

\bibitem{MeOHCr2O3}
{\O}. Borck and E. Schr\"oder,
J. Phys.: Cond. Matter \textbf{18}, 10751 (2006).

\bibitem{alkanes}
E. Londero, E.K. Karlson, M. Landahl, D. Ostrovskii, J.D. Rydberg, and E. Schr\"oder,
J. Phys.: Cond. Matter \textbf{24}, 424212 (2012).

\bibitem{phenol}
S.D. Chakarova-K\"ack, {\O}. Borck, E. Schr\"oder, and B.I. Lundqvist,
Phys. Rev. B \textbf{74}, 155402 (2006).

\bibitem{naphthalene}
S. D. Chakarova-K\"ack, E. Schr\"oder, B. I. Lundqvist, and D. C. Langreth,
Phys. Rev. Lett. \textbf{96} 146107 (2006).

\bibitem{PAHdimers}
S.D. Chakarova-K\"ack, A. Vojvodic, J. Kleis, P. Hyldgaard, and E. Schr\"oder,
New J. Phys. \textbf{12}, 013017 (2010).

\bibitem{chloroform}
J. {\AA}kesson, O. Sundborg, O. Wahlstr\"om, and E. Schr\"oder,
J. Chem. Phys. \textbf{137}, 174702 (2012).
 
\bibitem{adenine}
K. Berland, S.D. Chakarova-K\"ack, V.R. Cooper, D.C. Langreth, and 
E. Schr\"oder,
J. Phys.: Cond. Matt. \textbf{23}, 135001 (2011).

\bibitem{nucleobasesgraphene}
D. Le, A. Kara, E. Schr\"oder, P. Hyldgaard, and T.S. Rahman,
J. Phys.: Cond. Matter \textbf{24}, 424210 (2012). 

\bibitem{gpaw}
Open-source, grid-based PAW-method DFT code GPAW, see 
http://wiki.fysik.dtu.dk/gpaw/; J.J. Mortensen, L.B. Hansen, and K.W. Jacobsen,
Phys. Rev. B \textbf{71}, 035109 (2005).

\bibitem{soler}
G. Rom\'an-P\'erez and J.M. Soler,
Phys. Rev. Lett. \textbf{103}, 096102 (2009).

\bibitem{ASE}
Python-based atomic simulation environment, see
http://wiki.fysik.dtu.dk/ase/ 

\bibitem{Kintercalation}
E. Ziambaras, J. Kleis, E. Schr\"oder, and P. Hyldgaard,
Phys. Rev. B \textbf{76}, 155425 (2007).

\bibitem{fire} 
E. Bitzek, P. Koskinen, F. G\"ahler, M. Moseler, and
P. Gumbsch,
Phys. Rev. Lett. \textbf{97}, 170201 (2006).

\bibitem{morishige}
K. Morishige, K. Kawamura, and A. Kose,
J. Chem. Phys. \textbf{93}, 5267 (1990).

\bibitem{grimmePAHsize}
J. Antony and S. Grimme, 
Phys. Chem. Chem. Phys. \textbf{10}, 2722 (2008).
 
\bibitem{boyd}
S.L. Boyd and R.J. Boyd,
J. Chem. Theor. Comput. \textbf{3}, 54 (2007).

\bibitem{pettersson} J.B.C. Pettersson et al., unpublished.
 
\end{thebibliography}
\end{document}